

\input harvmac.tex
\input epsf

\def\figin{\epsfcheck\figin}\def\figins{\epsfcheck\figins}
\def\epsfcheck{\ifx\epsfbox\UnDeFiNeD
\message{(NO epsf.tex, FIGURES WILL BE IGNORED)}
\gdef\figin##1{\vskip2in}\gdef\figins##1{\hskip.5in}
instead
\else\message{(FIGURES WILL BE INCLUDED)}%
\gdef\figin##1{##1}\gdef\figins##1{##1}\fi}
\def\DefWarn#1{}
\def\figinsert{\goodbreak\midinsert}
\def\ifig#1#2#3{\DefWarn#1\xdef#1{fig.~\the\figno}
\writedef{#1\leftbracket fig.\noexpand~\the\figno}%
\figinsert\figin{\centerline{#3}}\medskip\centerline{\vbox{\baselineskip12pt
\advance\hsize by -1truein\noindent\footnotefont{\bf Fig.~\the\figno:}
#2}}
\bigskip\endinsert\global\advance\figno by1}

\lref\RZakh { V. Zakharov, V. L'vov and G. Falkovich,
``Kolmogorov spectra of turbulence I'', Springer-Verlag Berlin Heidelberg,
(1992)
}
\lref\RPol {A.Polyakov, unpublished}
\lref\RWil {J. Kogut and K. Wilson, {\sl Phys. Rep.} {\bf 12} (1974) 75}
\lref\RPok {A. Patashinski and V.Pokrovski, ``Fluctuation Theory
of Phase Transitions'', Pergamon Press, Oxford (1979)}
\lref\RConf { A. Polyakov, {\sl Nucl. Phys.} {\bf B396} (1993) 367}
\lref\RWyld {H. W. Wyld, {\sl Ann. Phys. } {\bf 14} (1961) 143}
\lref\RZawy {V.E. Zakharov and V.S. L'vov, {\sl Radiophysics and Quantum
Electronics} {\bf 18}(10) (1975) 1084}
\lref\RMalkin {V. Erofeev and V. Malkin, {\sl Zh.E.T.F.} {\bf 96}(5) (1989)
1666 [English translation: {\sl Sov. Phys. JETP} {\bf 69}(5) (1989) 943]  }
\font\title = cmr10 scaled \magstep 2
\font\subtitle = cmr10 scaled \magstep 1

{ \centerline{}
{ \bf April 1994 \hfill PUPT-1462 \break}
{ \null \hfill hep-th/9405077}
\vskip 30 pt
\centerline { \title  Probability Density, Diagrammatic Technique,}
\centerline {\title and Epsilon Expansion in the Theory of Wave Turbulence}
\baselineskip = 50 pt
\centerline {\subtitle V. Gurarie\footnote{${}^{\dagger}$}
{gurarie@puhep1.princeton.edu}}
}
\vskip 20 pt
\centerline {\subtitle \sl Department of Physics}
\centerline {\subtitle \sl Princeton University}
\centerline {\subtitle \sl Princeton, NJ 08544}
{\baselineskip=50 pt
\centerline {\subtitle \bf Abstract }}
\vskip 10pt

{ We apply the methods of Field Theory to study the turbulent regimes of
statistical systems. First we show how one can find their probability
densities. For the case of the theory of wave turbulence with four-wave
interaction we calculate them explicitly and study their properties.
Using those densities we show how one can in principle calculate any
correlation function in this theory by means of direct perturbative expansion
in powers of the interaction. Then we give the general form of the
corrections to the kinetic equation and develop an appropriate diagrammatic
technique. This technique, while resembling that of $\varphi^4$ theory,
has many new distinctive features. The role of the $\epsilon=d-4$ parameter
of $\varphi^4$ theory is played here by the parameter
$\kappa=\beta + d - \alpha - \gamma$ where $\beta$ is the dimension of the
interaction, $d$ is the space dimension, $\alpha$ is the dimension of the
energy spectrum and $\gamma$ is the ``classical'' wave density dimension.
If $\kappa > 0$ then the Kolmogorov index is exact, and if $\kappa < 0$
then we expect it to be modified by the interaction. For $\kappa$ a small
negative number, $\alpha<1$ and a special form of the interaction we compute
this modification explicitly with the additional assumption of the
irrelevance of the IR divergencies which still needs to be verified.
}

\vfill
\eject

\newsec {Introduction}

The theory of wave turbulence studies the stationary states of the
statistical classical (not quantum) system consisting of waves
with a small interaction.
(on the theory of wave turbulence see ref. \RZakh \ and references therein).
Its Hamiltonian can be written down in the following form
\eqn\Eham{ H = \sum_{p} \omega_{p} a^\dagger_{p} a_{p} + \sum_{p_1 p_2 p_3 p_4}
\lambda_{p_1 p_2 p_3 p_4} a^\dagger_{p_1} a^\dagger_{p_2} a_{p_3} a_{p_4}
}

It is just a collection of waves with the energy spectrum $\omega_{p}$ and
the four wave interaction $\lambda_{p_1 p_2 p_3 p_4}$ with the evident
properties $\lambda_{p_1 p_2 p_3 p_4} = \lambda_{p_2 p_1 p_3 p_4} =
\lambda_{p_3 p_4 p_1 p_2}$. Let us note that this Hamiltonian conserves
the total wave number
\eqn\Enumber {N=\sum_{p} a^\dagger_{p} a_{p}.}

The simplest stationary state of this system is a thermodynamic equilibrium
and its probability density is given by the well-known Gibbs
distribution $\exp ( - {H + \mu N \over T} )$. The basic
property of thermodynamic
equilibrium is that the detailed balance principle is satisfied. There are
as many waves going from the wave-number $p_1$ to $p_2$ as there are
ones going back. However there are other stationary states where that
principle is not satisfied, or in other words, while the total number
of waves coming to the given wave number $p$ is zero, there is a flux of waves
through the system. Which state will be chosen by the system depends on the
external conditions. If it interacts with a heat bath satisfying the
detailed balance principle, it will soon settle into the thermodynamic
equilibrium. If, on the other hand, the heat bath is so special that
it injects waves to the system at one wave number and removes them at
a different one, then the system will necessarily choose one of those
extra states. A simple example of the latter case is the waves on the
surface of water which are injected by, for example, the ship, at wave
lengths of the order of ship length, and are dissipated at much smaller
lengths by viscosity.

The theory of turbulence concerns itself with studying the ``flux''
states as the thermodynamic equilibrium has already been studied in
great details by Statistical Physics. The standard hydrodynamics
turbulence is also the example of the flux states since
here we have a very complicated motion of liquid with a stationary
probability distribution which is characterized by the flux of
energy or other conserved quantities through various scales of
vertex motion.

While the existence of those extra states was known for a long time
and they were analyzed with the aid of kinetic equations (ref. \RZakh ), the
explicit form of the probability distribution was never found.
But let us remember that the Liouville theorem of the statistical mechanics
tells us that the probability distribution of any system should be an
additive integral of motion. The next step is usually to assume the
system doesn't have any additive integrals of motion different from
the energy and the particle (or wave) number. That's how Gibbs distribution
is obtained. But in fact any dynamical system has as many integrals of
motion as the number of its degrees of freedom. Really, any initial
conditions expressed in terms of the changing variables give us a formal
integral of motion. The important role such integrals play in the
description of the flux-states of the statistical systems
 was first noticed by A. Polyakov
 (ref. \RPol).
Essentially such probability distributions mean that we prepared first
a statistical  ensemble with the probability $\rho(q,p)$ where $q$ and $p$
 are the
initial coordinates and momenta respectively. Then we follow its time
development. A correlation function of some quantity $X(q,p)$ at the moment
$t=0$ is given by $$ \int dq dp X(q,p) \rho(q,p).$$ To find this function
at a later time $t$ we need to use the solutions of  the equations of motion,
$p(t)$ and $q(t)$, to express the initial conditions in terms of the values
of the coordinates and the momenta at a later time $t$, $q(q(t),p(t),t)$
and $p(q(t), p(t),t)$. The correlation function will then be given by
$$ \int dq dp X(q(t),p(t)) \rho(q,p) = \int dq(t) dp(t) X(q(t),p(t))
\rho^\prime $$
where $\rho^\prime$ is the same probability distribution expressed
in terms of the current coordinates and momenta $q(t),p(t)$. The Liouville
theorem of classical mechanics $dp dq = dp(t) dq(t) $ allowed us to make that
change of variables.

Constructed in this way, $\rho^\prime$ is not a stationary probability
distribution. But we expect
it to have some limit as $t \rightarrow \infty$ otherwise the system
we study does not  have  stationary states.  But even if such a limit
exists, it does not automatically imply all the correlation functions
will also have some well defined limit as $t \rightarrow \infty$.
This phenomenon will be discussed later in this paper. Now it
will be
 enough to say that the additional conditions $\rho^\prime$ should satisfy
to give a stationary state for the system is called the kinetic equation.
To uniquely find the probability distribution one must solve the kinetic
equation. Thermodynamic equilibrium is often just one of many solutions.

The main problem of the theory of turbulence is to solve the kinetic equation
and to find the correlation functions, especially the two point correlation
function $<a^\dagger_p a_p>$ which gives the spectrum of the theory.

In this paper we are going to study the kinetic equation for the system \Eham \
and develop a diagrammatic technique to represent the infinite series
in powers of $\lambda$  we will have to deal with. This technique
while somewhat similar to that of $\varphi^4$ theory will have many
features
not encountered in Field Theory before. It may be even viewed as
 a generalization
of $\varphi^4$ technique  since in the limit of thermodynamic equilibrium
it reduces
to the standard $\varphi^4$ theory.

Nevertheless we are going to find
some specific cases, namely when the interaction coefficient $\lambda$ is
just a product of the external momenta, when  one can extract the infinite
series of most divergent diagrams, just like in $\varphi^4$ theory in four
dimensions, with the condition of the dimensionlessness of the interaction
coefficient, of course.
With some additional assumptions, in particular neglecting the
IR divergencies, we shall make use of it to
calculate the correction to the ``classical'' spectrum of the theory.

Technically there is no need to know the probability distribution to
compute all the correlation functions of the theory \Eham \ as discussed
later in the paper. But we hope that the study of the probability
distribution for \Eham \ will allow us to understand better the
behavior of the correlation function of the hydrodynamic turbulence
where no reliable methods of finding the correlation functions exist. We
shall discuss that in the end of the paper.

The method which was widely used to compute the correlation functions
thus far is called the Wyld's technique (see refs. \RWyld,
\RZawy). It is essentially a classical nonequilibrium diagrammatic technique,
and it deals with the development of the correlation functions in time
unlike the technique
proposed in this paper which deals with the equal time correlation
functions, as in statistical mechanics.
Recently a paper \RMalkin \  was published where the authors claimed they
could in principle compute the kinetic equation up to any order of
the interaction using the Wyld's technique. We think that  while
the kinetic equation obtained in this paper should be equivalent
with that of \RMalkin,
our method is  easier for practical calculation.

\newsec {The Probability Distribution}

According to the program outlined in the Introduction, first of all
we need to
study all possible integrals of motion for the system \Eham.
Let  us see how one can construct those integrals.
We shall start by assigning the variables $a_p$ and $a^\dagger_p$ the
initial values $a_{p}(t=0)=a^0_p$, $a^{\dagger}_p(t=0)=a^{\dagger 0}_p$.
Then by solving the equations of motion one can find $a(t)$ and $a^\dagger(t)$
as functions of $a^0$ and $a^{\dagger 0}$ and time. By inverting those
functions, one can find $a^0$ and $a^{\dagger 0}$ as functions of time,
$a(t)$ and $a^\dagger (t)$. But by definition $a^0$ and
$a^{\dagger 0}$ do not depend on time. So they are the integrals of motion.
If we want to construct something out of them which can play the role of
the density of waves, we should consider the linear combination
\eqn\EFzero {F=\sum_{p} f_p a^{\dagger 0}_{p} a^0_p }
where $f_p$ are some arbitrary coefficients.
After being expressed in terms of $a(t)$, $a^\dagger (t)$ and $t$ it becomes
a valid integral of motion. Its explicit time dependence can be
eliminated by passing to the limit $t \rightarrow \infty$.

In order to find the  explicit form for $F$
we need to solve the equations of motion.
The Hamiltonian
\Eham \ allows us to solve them perturbatively and we can obtain $F$ in terms
of a series in powers of $\lambda$. We can even avoid solving the equations
of motion if we use the following procedure. One should look for $F$ in terms
of a power series
\eqn\EFf { \eqalign {F=\sum_{p} f_p a^\dagger_p a_p +
\sum & \Lambda_{p_1 p_2 p_3 p_4}
a^\dagger_{p_1} a^\dagger_{p_2} a_{p_3} a_{p_4} +
\cr + & \sum \Omega_{p_1 p_2 p_3
p_4 p_5 p_6} a^\dagger_{p_1} a^\dagger_{p_2} a^\dagger_{p_3} a_{p_4} a_{p_5}
a_{p_6} + \dots.  } }
Here $\Lambda$ and $\Omega$ are some still unknown functions. We impose the
condition on $F$ that it is an integral of motion, or $\{ HF \} =0$, $\{$
and $\}$ being the Poisson brackets.
It allows us to find those functions to get
\eqn\Elambda{ \Lambda_{p_1 p_2 p_3 p_4} = {f_{p_1} + f_{p_2} - f_{p_3} -
f_{p_4} \over
\omega_{p_1} + \omega_{p_2} - \omega_{p_3} - \omega_{p_4} - i \epsilon }
\lambda_{p_1 p_2 p_3 p_4}}
\eqn\Eomega{\Omega_{{p_1} {p_2} {p_3} {p_4} {p_5} {p_6}} = 4 \sum_{{p_7}}
{ ( \lambda_{{p_7} {p_1} {p_5} {p_6}} \Lambda_{{p_2} {p_3} {p_4} {p_7}} -
\Lambda_{{p_7} {p_1} {p_5} {p_6}} \lambda_{{p_2} {p_3} {p_4} {p_7}} )
\over \omega_{p_1} + \omega_{p_2} + \omega_{p_3}
- \omega_{p_4} - \omega_{p_5} -
\omega_{p_6} - 2 i \epsilon} .}
We can in principle
find recursively all the terms in the series \EFf, one after another.

A few words must be said about $\epsilon$'s which appear in the denominators.
Technically, when we compute Poisson brackets no $\epsilon$'s appear. But
we must introduce them to avoid the poles in \Elambda \ and \Eomega \
in order for the sums in \EFf \ to make sense. This procedure is
legitimate only if we deal with the infinite number of degrees of
freedom, that is the sums in \EFf \ are really the integrals. If we
had dealt
with finite dimensional system, we would have obtained unavoidable zeros in the
denominators of \EFf \ and the integrals of motion we discuss would not
have
existed. This is a very important observation.

Another way to see why $\epsilon$
must appear is to use the original procedure of finding $F$. One would
discover that the function $F$ depends on the expressions like $(1- \exp (-
i \omega
 t)) / \omega$ and the only way to pass to the limit $t \rightarrow \infty$
is to assume that $\omega$ has an infinitesimal imaginary part. The last
procedure ensures the signs of $\epsilon$'s are correct.

It may seem somewhat strange that we paid attention to writing
 $2 i \epsilon$ in the denominator
of $\Omega$ as $1/(x-i \epsilon)=1/x + i \pi \delta(x)$ regardless
of the magnitude of $\epsilon$. However, later we shall see
the cases when the magnitude of $\epsilon$ becomes important, in particular
when we deal with the expressions like $\epsilon \delta(x)/(x-i \epsilon)$.
To check the magnitude of $\epsilon$ of the denominator of $\Omega$, one
should follow the finite time procedure described in the previous
paragraph carefully. One would discover that the denominator of $\Omega$
came from summing up the denominators of two $\Lambda$'s included in the
definition of $\Omega$. If each of these $\Lambda$'s has $- i \epsilon$
in its denominator, then there should be $- 2 i \epsilon$ in the
denominator of $\Omega$. This point will be clarified later in the paper.

The parameter $\epsilon$ plays the role of a regulator in the theory
to be discussed and in this sense is analogous to the cutoff used in
Quantum Field Theory\footnote{${}^{\dagger}$} {It
is a time cutoff. The momentum
cutoff  will also have to be imposed if needed.
}.
The function
$F$ by itself is only an approximate integral of motion as long as $\epsilon$
is finite and we must pass to the limit $\epsilon \rightarrow 0$ in
all our calculations.

The construction of the integrals of motion \EFf \ represents a certain
paradox. The system given by \Eham \ is not necessarily integrable.
Moreover, we are not interested in integrable systems. If the system
were integrable, we could always make a change of variables to
bring it to the form of noninteracting oscillators where no fluxes
of energy could exist.
Nevertheless we have just constructed the complete set of the integrals
of motion for \Eham. Something must be wrong with those integrals,
otherwise the system \Eham \ would be integrable.

It is those poles in the denominators of \Elambda \ and \Eomega \
 which are responsible for those integrals not to give rise to any
integrability. They actually make the integrals \EFf \ to be multi-valued
functions which cannot be used to construct those tori the trajectories
of integrable systems lie on. But \EFf \  can perfectly be used to
play the role of probability distribution if one also introduces
$\epsilon$'s to regularize them. The finite time computation, as was
discussed above, validates the introduction of $\epsilon$'s in the
 form as in \Elambda \ and \Eomega.

Now our statement is  all the statistical states of the theory \Eham \ can
be described by the probability density given by
\eqn\Eprob { \exp (-F).}
Even a thermodynamic equilibrium is a particular case of \Eprob, with $f_p=
\omega_p$
or $f_p=1$. All the correlation functions of the theory can be found
using
\eqn\Ecorr{ <X> \equiv {1 \over Z}
\int \prod_{p} da_{p} \prod_{p} da^\dagger_{p} X \exp( -F) }
 where $Z$ is a partition function
\eqn\epart{ Z= \int \prod_{p} da_{p} \prod_{p} da^\dagger_{p}
\exp( -F) }
In a complete analogy with Quantum Field Theory, the
role of the partition function in the denominator amounts just
to removing ``vacuum diagrams'', or contractions totally inside the
interaction,
so we will usually omit it in our formulae while keeping that in mind.

We are going now to demonstrate how one can derive a well-known kinetic
equation (in the first nonvanishing order of perturbation theory)
using our probability distribution. The kinetic equation
is just a statement that the derivative with respect to the time of the
correlation function $<a^\dagger_p a_p>$ is zero. In other words,
\eqn\Ekinn{
\eqalign{< {d (a^\dagger_p a_p) \over dt}> &= < \{H, a^\dagger_p a_p\} > =
 \cr
= - i \sum_{p_1 p_2 p_3 p_4}
\lambda_{p_1 p_2 p_3 p_4} & < a^\dagger_{p_1} a^\dagger_{p_2} a_{p_3} a_{p_4}>
(\delta_{p p_1} + \delta_{p p_2} - \delta_{p p_3} - \delta_{p p_4}) =0}
.}

Since the function $F$ is real, one can easily see from \Ekinn \ that only
the imaginary part of the correlation function $<a^\dagger_{p_1}
a^\dagger_{p_2} a_{p_3} a_{p_4} >$ will contribute to the kinetic equation
which can actually written down in the form
\eqn\Ekinnexact { - 4  i \sum_{p_2 p_3 p_4} \lambda_{p p_2 p_3 p_4}
{\rm Im} <a^\dagger_{p} a^\dagger_{p_2} a_{p_3}
a_{p_4}> = 0 }

Let us find the four point correlation function perturbatively. Its
zero order value is equal to
$$\eqalign {
<a^\dagger_{p_1}  a^\dagger_{p_2} a_{p_3} a_{p_4} >
\equiv {1 \over Z} \int
\prod_{p} da_{p} & \prod_{p} da^\dagger_{p} a^\dagger_{p_1} a^\dagger_{p_2}
a_{p_3} a_{p_4} \exp( - \sum f_p a^\dagger_p a_p) = \cr
& ={1 \over f_{p_1} f_{p_2}}
(\delta_{p_1 p_3} \delta_{p_2 p_4} + \delta_{p_1 p_4} \delta_{p_2 p_3})}
. $$
This expression is real so it will not contribute to the equation \Ekinnexact.
Now the first order contribution to that function is obtained by
expanding the exponent in powers of $\lambda$ and is given by
\eqn\Eint { - 4\lambda_{p_1 p_2 p_3 p_4} {f_{p_3}+f_{p_4}-f_{p_1}-f_{p_2} \over
\omega_{p_3}+\omega_{p_4}-\omega_{p_1}-\omega_{p_2} - i \epsilon}
{1 \over f_{p_1} f_{p_2} f_{p_3} f_{p_4}} + R  }
where $R$ is a real expression coming from the contractions
inside the interaction which is not interesting because it is
real and also because it is zero in case if the interaction $\lambda$
conserves the momentum.
Imaginary part of \Eint \ is
$$  4 i \pi \lambda_{p_1 p_2 p_3 p_4} (f_{p_1} + f_{p_2}
- f_{p_3} - f_{p_4}) \delta (\omega_{p_1} + \omega_{p_2} - \omega_{p_3} -
\omega_{p_4}) {1 \over f_{p_1} f_{p_2} f_{p_3} f_{p_4} }$$
and after substituting it to \Ekinnexact \ we arrive at the kinetic
equation
\eqn\Ekin{
 16 \pi \sum_{p_2 p_3 p_4} \lambda^2_{p p_2 p_3 p_4} \delta (\omega_p -
\omega_{p_2} - \omega_{p_3} - \omega_{p_4}){ f_p+ f_{p_2}-f_{p_3}-f_{p_4}
\over f_{p} f_{p_2} f_{p_3} f_{p_4} } = 0 .}

One can easily recognize the standard kinetic equation of ref. \RZakh \
if one substitutes with the same accuracy
$<a^\dagger_p a_p>=1/f_p$.

\newsec {The Anomaly}

As it has already been mentioned in the Introduction,
it is highly nontrivial that the kinetic equation we obtained
is not identically zero. How can a stationary probability distribution
give rise to a correlation function depending on time? More precisely
we can use the following argument.
For any quantity $X$
$$< {dX \over dt}>=<\{HX\}>=\int \{HX\} \exp (-F)=\int \{HF\} X
\exp (-F)\equiv 0 $$
where we took the integral by parts and used that $F$ is the
integral of motion, that is $\{HF\}=0$. But it contradicts  our
computations. The resolution of this paradox lies in the fact that
$F$ is the integral of motion only in the limit $\epsilon \rightarrow
0$. If we compute $\{H F\}$ we shall discover that it is of the order of
$\epsilon$. But then, when computing the integral $\int \{H F\} X \exp
(-F)$, we shall find that it in itself is of the order of $1/ \epsilon$.
Those epsilons cancel and we can safely pass to the limit $\epsilon
\rightarrow 0$ to get the finite answer for the kinetic equation.
The effect is completely analogous to the anomaly of
Quantum Field Theory. By imposing the cutoff $\epsilon$ on the theory
we violate the conservation of $ F$ and it  is never fully restored
after removing the cutoff\footnote{${}^{\dagger}$}{I am
indebted to I. Kogan for pointing out
this correspondence.}.

It is instructive to rederive the kinetic equation in this fashion.
The Poisson bracket $\{HF\}$ can be recursively computed to give
\eqn\Epois{ \{HF\} =-\epsilon \Lambda_{{p_1} {p_2} {p_3} {p_4}} a^\dagger_{p_1}
a^\dagger_{p_2}
a_{p_3} a_{p_4} - 2 \epsilon \Omega_{{p_1} {p_2} {p_3} {p_4} {p_5} {p_6}}
a^\dagger_{p_1} a^\dagger_{p_2} a^\dagger_{p_3}
a_{p_4} a_{p_5} a_{p_6} - \dots . }
Then
\eqn\Epart { < {d (a^\dagger_p a_p) \over dt} > = \int a^\dagger_p a_p \{HF\}
\exp (
-F).}
The first order contribution to this expression is
\eqn\Eloop{ \eqalign {
\int \sum_{{p_1} {p_2} {p_3} {p_4}} a^\dagger_p a_p a^\dagger_{p_1}
a^\dagger_{p_2} a_{p_3} a_{p_4} & \exp ( \sum_k - f_k
a^\dagger_k a_k) \cr & {f_{p_1}+f_{p_2}-f_{p_3}-f_{p_4} \over \omega_{p_1} +
\omega_{p_2} - \omega_{p_3} -
\omega_{p_4} - i \epsilon} \lambda_{{p_1} {p_2} {p_3} {p_4}} \epsilon =0.}}
It is equal to zero even with $\epsilon > 0$ as the contractions will
enforce $f_{p_1}+f_{p_2}-f_{p_3}-f_{p_4}=0$.

The second order contribution to \Epart \ is given by
\eqn\Eterrible {\eqalign{ & \int a^\dagger_p a_p
(- 2 \epsilon \Omega_{{p_1} {p_2} {p_3} {p_4} {p_5}
{p_6}} a^\dagger_{p_1}
a^\dagger_{p_2} a^\dagger_{p_3} a_{p_4} a_{p_5} a_{p_6} + \cr + \epsilon
 \Lambda_{{p_1} {p_2} {p_3} {p_4}}&
\Lambda_{{p_5} {p_6} {p_7} {p_8}} a^\dagger_{p_1} a^\dagger_{p_2}
a_{p_{p_3}} a_{p_4} a^\dagger_{p_5} a^\dagger_{p_6}
a_{p_7} a_{p_8})   \exp ( \sum_k - f_k a^\dagger_k a_k).}}
The integral involving $\Omega$ will be of the order of $1/ \epsilon$ as
the contractions will ensure the denominator in \Eomega \ to be equal to $- 2 i
\epsilon$. After rather long and tedious computations (which  actually
become quite simple with some practice) we obtain
\eqn\Eentropy { 2 \epsilon \int a^\dagger_p a_p \Omega a^\dagger a^\dagger
a^\dagger a a a
\exp(-\sum f_k a^\dagger_k a_k) = Q_p+P_p - K_p } where
\eqn\Ep { \eqalign {P_p=\sum_{ p_1 p_2 p_3 p_4}
{ 4 \pi \over f_p f_{p_1} f_{p_2} f_{p_3}
f_{p_4} } & \lambda^2_{p_1 p_2 p_3 p_4} \delta (\omega_{p_1} + \omega_{p_2} -
\omega_{p_3} - \omega_{p_4}) \cr & (f_{p_1}+f_{p_2}-f_{p_3}-f_{p_4})^2
[ \delta_{p p_1} +\delta_{p p_2}
+\delta_{p p_3}+\delta_{p p_4}],}  }
\eqn\Eq { \eqalign {Q_p=
\sum_{ p_1 p_2 p_3 p_4} &
{ 4 \pi \over f_p f_{p_1} f_{p_2} f_{p_3}
f_{p_4} } \lambda^2_{p_1 p_2 p_3 p_4} \cr &
\delta (\omega_{p_1} + \omega_{p_2} -
\omega_{p_3} - \omega_{p_4}) (f_{p_1}+f_{p_2}-f_{p_3}-f_{p_4})^2,
}}
and
\eqn\Ek { K_p=
 \sum_{p_2 p_3 p_4} {16 \pi \over
f_p f_{p_2} f_{p_3} f_{p_4} }
\lambda^2_{
p p_2 p_3 p_4} (f_p + f_2 - f_3 - f_4) \delta (\omega_p + \omega_{p_2} -
\omega_{p_3} - \omega_{p_4} ) .}
The term $Q_p$ in \Eentropy \ came from contracting of $a_p$ with
$a^\dagger_p$ while the rest of the terms came from intercontracting
those
operators with the interaction.
We recognize the kinetic equation \Ekin \ in $K_p=0$ while the term
$P_p$ has the meaning of the entropy of waves\footnote{${}^{\dagger}$}
{See ref. \RZakh \ for details on the entropy definition.}.

As for the integral involving $\Lambda^2$, its divergent part is
determined by the formula $$ {1 \over (x + i \epsilon) (x - i \epsilon)
} \rightarrow {\pi \over \epsilon} \delta ( x), \ \ \epsilon \rightarrow 0 $$
to give
\eqn\Eterri { \epsilon \int a^\dagger_p a_p \Lambda a^\dagger a^\dagger a a
\Lambda
a^\dagger a^\dagger a a \exp ( - f a^\dagger a) =
Q_p+P_p. }
The difference between \Eentropy \  and \Eterri \ gives us exactly the kinetic
equation \Ekin \  as we expected.

However the calculations we just completed imply more than just
checking the validity of integrating by parts. We can use the above
formulae to calculate the corrections to the wave density $<a^\dagger_p
a_p>$. We obtain
\eqn\Etwop {<a^\dagger_p a_p > = {1 \over f_p} + \int a^\dagger_p a_p (- \Omega
a^\dagger a^\dagger a^\dagger a a a + {\Lambda \Lambda \over 2}
a^\dagger a^\dagger a a a^\dagger a^\dagger a a ) \exp( -f a^\dagger a)
.}
These integrals have already been computed. The only difference with the
integrals in \Eentropy \  and \Eterri \
is the absence of the ``vacuum diagrams''
or $Q_p$, and of course $\epsilon$ will not be canceled.  That makes
a crucial difference with \Eterrible. Not only the expressions we shall
obtain will be divergent with $\epsilon \rightarrow 0$, but also we need
to extract the final parts from them, for example by using
\eqn\Eexpans{ {1 \over (x + i \epsilon) (x - i \epsilon)
} \rightarrow {\pi \over \epsilon} \delta ( x)- {1 \over x} {\partial \over
\partial x} + o(\epsilon)}
We shall discuss the formulae like \Eexpans \ later in more details.
Now we just note that in addition to the divergent with $\epsilon
 \rightarrow 0$  part, there will be a finite part which we shall not
calculate here, leaving it to the discussion of the self energy corrections
later in this paper.

 The  answer will be
\eqn\Eformal { <a^\dagger_p a_p> = {1 \over f_p} - {1 \over 2 \epsilon}
( P_p-K_p) + {1 \over 2 \epsilon} P_p + {1 \over \epsilon} O(\epsilon)
={1 \over f_p} + {1 \over 2 \epsilon
}K_p + {\rm finite \ \ part}}

We see that the second order contribution to the wave density
is clearly divergent and it should not be strange. The $\epsilon$
regularization actually plays the role of cutting off the times greater
than $1 / \epsilon$ and if something is equal to $1 / \epsilon$, that
means it is divergent with time going to infinity which can happen to
the wave density if the kinetic equation is not satisfied. However if
it is satisfied, we would rather expect no divergencies in \Eformal.
 So it is reasonable to believe that the divergent
higher order corrections to \Eformal \
will sum up to be $1/\epsilon$  multiplied by  the
exact kinetic equation \Ekinnexact. As long as the kinetic equation is
satisfied, we can neglect those corrections.

\newsec { Lowest Order Corrections to the Kinetic Equation}

With the experience of calculating different correlation functions we
acquired in the previous section, we are ready to start computing
the corrections
to the kinetic equation. All we need to do is to take \Ekinnexact \ and compute
the imaginary part of the four point correlation function order by order
by expanding $\exp(-F)$. The calculations are messy, but pretty
straightforward. In this section we are are going to discuss some of the lowest
order corrections to the equation \Ekin, the ones which are not the corrections
to the propagators, just to get acquainted with them.
They consist
of six terms. We begin with considering the first three of them. They are
\eqn\Epuzyr { \eqalign{\sum_{p_1 p_2 p_3 p_4 p_5 p_6 }8
\pi \lambda_{p_1 p_2 p_3 p_4}
\lambda_{p_3 p_4 p_5 p_6} \lambda_{p_5 p_6
p_1 p_2} & \biggl( {f_{p_1}+f_{p_2}-f_{p_3}-f_{p_4} \over f_{p_1} f_{p_2}
f_{p_3} f_{p_4} } \times
\cr \delta (\omega_{p_1} + \omega_{p_2} - \omega_{p_3} -
\omega_{p_4}) & {f_{p_5}+f_{p_6} \over f_{p_5} f_{p_6} (\omega_{p_5} +
\omega_{p_6} - \omega_{p_3} - \omega_{p_4})} + \cr +
(3,4) \leftrightarrow (5,6) +  (1,2) \leftrightarrow (5,6) \biggl)
 & [\delta_{p p_1} + \delta_{p p_2} - \delta_{p p_3} - \delta_{p p_4}]
 } }
$(1,2) \leftrightarrow (5,6)$ means that this expression differs from
the explicitly written one by exchanging the indeces.
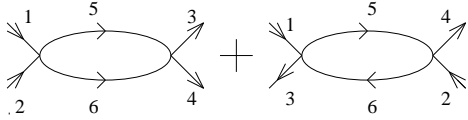
\ifig\Fdiag{Some of the lowest order corrections to the four point function}
{\epsfxsize
2.0in\epsfbox{diag.ps}}

It is easy to represent
the way how $\lambda$'s are connected with each other
by a simple diagram borrowed from the $\varphi^4$ theory. The only
feature we have to add is the little arrows on each line representing
whether the line goes into the vertex or from the vertex. Since
$\lambda_{p_1 p_2 p_3 p_4}$ is included in any correction (it comes
from the exact equation \Ekinnexact \ rather than from the perturbative
expansion), there is no need to represent it. The two remainings $\lambda$'s
are represented by the first of the diagrams on \Fdiag.

As for the coefficient in front of the $\lambda$'s, it doesn't resemble
anything in $\varphi^4$ theory.
 The only
familiar feature it retained from that theory
is the product of all the propagators $1/(f_{p_1} f_{p_2} f_{p_3}
f_{p_4} f_{p_5} f_{p_6})$. All the rest came from the interaction
coefficients \Eomega \ and \Elambda  \ which are very nontrivial.
The next section of this paper will be devoted to their analysis.
 Now we shall only note  the fact that the
first term in \Epuzyr \  coincides with the kinetic
equation \Ekin \ itself multiplied by
\eqn\Evstav {{f_{p_5}+f_{p_6} \over f_{p_5} f_{p_6} (\omega_{p_5} +
\omega_{p_6} - \omega_{p_3} - \omega_{p_4})}. }
The two intermediate lines in the diagram should be represented by \Evstav.
The second and the third terms in \Epuzyr \ are constructed from the first one,
so we don't need to discuss them separately.

The expression \Evstav \ has a
simple physical meaning. The scattering represented
by the diagram we consider can happen in two ways. Either the waves
1 and 2 scatter to 5 and 6, and then 5 and 6 scatter to 3 and 4, or
the waves 5 and 6 first scatter to 3 and 4, and only then  1 and 2
scatter to 5 and 6. The first amplitude should be proportional to $(n_5+
1)(n_6+1)$ where $n=1/f$ is a wave density while the second to
$n_5 n_6$. The ``energy denominators'', represented by the $\omega$'s
which are of the same origin as the similar denominators in quantum
mechanical perturbation theory, will have opposite signs in these two cases
so the amplitude will be proportional to $(n_5+1)(n_6+1)-n_5 n_6 \approx
n_5+n_6$, that is exactly \Evstav.

The three terms corresponding to the second diagram from \Fdiag \
can be computed to be
\eqn\antpuz { \eqalign{\sum_{p_2 p_3 p_4 p_5 p_6 }32
\pi \lambda_{p_1 p_2 p_3 p_4}
\lambda_{p_4 p_5 p_2 p_6} \lambda_{p_6 p_3
p_1 p_5} &  \biggl( {f_{p_1}+f_{p_2}-f_{p_3}-f_{p_4} \over f_{p_1} f_{p_2}
f_{p_3} f_{p_4} } \times
\cr \delta (\omega_{p_1} + \omega_{p_2} - \omega_{p_3} -
\omega_{p_4})& {f_{p_5}-f_{p_6} \over f_{p_5} f_{p_6} (\omega_{p_5} +
\omega_{p_4} - \omega_{p_6} - \omega_{p_2})} +\cr +
(3,1) \leftrightarrow (5,6) + (2,4) \leftrightarrow (5,6) \biggl)
& [\delta_{p p_1} + \delta_{p p_2} - \delta_{p p_3} - \delta_{p p_4}]
 } }
Here the intermediate waves propagate in the opposite directions, so
the amplitude is proportional to $(n_5+1)n_6-n_5(n_6+1) \approx n_6-n_5$.

\newsec {Diagrammatic Technique}

We can proceed with calculating higher order diagrams in the same fashion.
But first we shall consider another approach of calculating the
corrections to the four point correlation function, the one which
doesn't require the knowledge of the probability distribution.
While it gives the same results, it is sometimes simpler to use for
practical computations. This approach  is based on the
idea that the correlation functions should not depend on time.
For example, if we compute the time derivative of the four point
correlation function, we obtain
\eqn\Etime{ \eqalign {
<\{ H, a^\dagger_{1}  a^\dagger_{2} a_{3} a_4 \}> =  i < & a^\dagger_1
a^\dagger_2 a_3 a_4 > (\omega_1+\omega_2-\omega_3-\omega_4) + \cr +
\sum_{p_1 p_2 p_3 p_4} & < \{ \lambda_{p_1 p_2 p_3 p_4} a^\dagger_{p_{1}}
a^\dagger_{p_{2}} a_{p_3} a_{p_4} , a^\dagger_{1}  a^\dagger_{2} a_{3}
a_4 \} > = 0 }}
which expresses the four point correlation function in terms of the
six point one. Clearly,
 the zero order value of the six
point correlation function will give us the first order value to the
four point one and so on. Quite analogously, to obtain the first order
correction to the six point correlation function, we write that its
derivative is zero to express it in terms of eight point correlation
function and so on. In order to avoid the poles while using \Etime \  we
replace $\sum \omega$ by $\sum \omega + i \epsilon$. The sign
of the $\epsilon$ can be checked to be $+$ by computing everything at
finite time\footnote{${}^{\dagger}$} {That means  by solving the
equation for the time derivative of the four point function and choosing
the imaginary part for the $\omega$ to be such that there's a finite
limit at
$t \rightarrow \infty$.}.

Our aim is to find the perturbation series for the four point correlation
function, as clear from the exact kinetic equation \Ekinnexact.
{}From the above
consideration we see that the n-th order correction will be of the form

\eqn\Ebasis {i^n < a^\dagger_1 a^\dagger_2 a_3 a_4>_{n}=<{ \{H_i, { \{H_i,
\dots, { \{ H_i, a^\dagger_1
a^\dagger_2 a_3 a_4 \} \over \sum (\omega) + i \epsilon} \} \over
\sum (\omega) + (n-1) i \epsilon} \} \over
\sum (\omega) + n i \epsilon} >_{0} }
where $H_i$ is the interaction part of the hamiltonian \Eham, $$H_i=
\sum_{p_1 p_2 p_3 p_4} \lambda_{p_1 p_2 p_3 p_4}
a^\dagger_{p_1} a^\dagger_{p_2} a_{p_3} a_{p_4}, $$
and
we retain just the zero order value for the correlation function on the
 right hand side of \Ebasis.

Let us see how \Ebasis \ works. First we compute the Poisson bracket
$$\{ H_i, a^\dagger_1
a^\dagger_2 a_3 a_4 \}=i \sum \lambda_{p_1 p_2 p_3 1} a^\dagger_{p_1}
a^\dagger_{p_2} a_{p_3} a^\dagger_{2} a_3 a_4 + \dots$$
We represent the product $a^\dagger_1 a^\dagger_2 a_3 a_4$ by a vertex
with the incoming lines 1 and 2 and the outgoing lines 3 and 4. Taking
the Poisson bracket means attaching another vertex to the original one
 with the
lines $p_1$, $p_2$, $p_3$, and $1$, the last line going to the first vertex.
Then, according to \Ebasis \
 we divide all we obtained by $\omega_1+\omega_2
-\omega_{3}-\omega_4+
i \epsilon$. Graphically, it is just the sum and difference of $\omega$'s
corresponding to the incoming and the outgoing lines of the first vertex.
Then we continue the procedure. Taking the Poisson bracket again means
attaching another vertex to the diagram we are building, and then
we need to divide by the sum and difference of $\omega$'s corresponding
to the first and the second vertex taken together. At last, when we arrive
at the $n$th order picture, which  is called the tree diagram,
 we need to compute the Gaussian correlation function, that means
to close each outgoing external line of the tree with the incoming line
and multiplying the expression by the product of the propagators $1/f$
for each Gaussian contraction. Simultaneously our picture transforms itself
into something resembling the standard $\varphi^4$ theory diagram.
The distinctive feature is the arrows on the lines
(to distinguish between the incoming and outgoing
lines) as we already saw in the previous section. Also, if we do everything
honestly, we need to close the external lines 1, 2, 3, and 4 into one
additional vertex, but that makes the drawing rather clumsy, so we
shall be using the standard diagrams like those shown on \Fdiag.


In short, we see that the right hand side of \Ebasis \
is constructed by connecting $\lambda$'s from different $H_i$
together by means of the
Poisson brackets to form a tree diagram and multiplying $\sum \omega$
in the denominator along the way, then by pairing all
$a$ with $a^\dagger$ to form a real diagram writing the propagators
$1/f$ for each line which was formed by pairing and not by a Poisson
bracket connections. The symbol $\sum \omega$ means first we take
the sum  of $\pm \omega$'s belonging to the incoming or
outgoing lines of the first vertex, then we take the sum
of $\pm \omega$ belonging to the incoming or outgoing lines of the
first and the second vertex along the tree ``trunk'' combined and so on.

 We can reverse the procedure to
say that first we should draw a $\varphi^4$ diagram, then cut some
of its lines to turn it into a tree diagram and write down the
product of $\sum \omega$ taken along the tree trunk. Then we
multiply what we got by the product of $f$ taken alone the trunk and
divide it by the product of all $f$'s. And the last step should be
summing over all different trees which can be curled up to form
the same diagram.  Each tree should have a direction along which
we go and each trunk line then has a sign depending on whether
this line has an arrow alone or opposite to that general direction.
The sign of the diagram is the product of the signs of the
trunk lines. It follows from the property of the Poisson bracket
to have opposite signs in case when $a$ is connected to $a^\dagger$
or vice versa.

We can go a little bit further in the formulation of our rules of
reading diagrams.
Namely if the tree have many branches, there are many ways we can
sum up the $\omega$'s. We can go along the first branch of the tree,
then the second, or first along the second, then the first, or
we can go up half of the first branch, then along the second, and
then finish off the first one. And we need to take the sum of all
the expressions involving $\omega$'s we obtain in this way.

To compute this sum is a difficult job, but fortunately the expressions
conspire in such a way as to give rise to the formula which can be
written down using the Feynman rules given below which work for
both the trees with branches and without, and so they give the general
way of computing any diagram without doing the actual computation.

\centerline {The Feynman rules. }

In order
to write down the expression corresponding to a given diagram, we need:

\noindent
1) To write down the product of $\lambda$'s corresponding to the vertices
of the diagram, just like in $\varphi^4$ theory.

\noindent
2) To write down the product of all the propagators $1/f$ corresponding
to each line of the diagram, again just like in $\varphi^4$ theory.

\noindent
3) And to compute a special prefactor.
First we need to make up a list of
all possible expressions $\sum \omega$ taken around each vertex
of the diagram and around each combination of vertices. Then we need
to cut some of the lines of the diagram to convert it into a tree
diagram. The tree diagram by definition always starts from the external line
while the external lines are thought as being connected together
by some ``external"
vertex. The expression corresponding to the tree will be
 the product of all $f$'s taken
along the trunk and the product of as many $1/(\sum \omega)$
taken from the list we made up
as there are
$f$'s, each of the $\sum \omega$ having one and only
one particular omega belonging to one of the trunk lines. One can show
there is one and only one subset of $\sum \omega$ with this property
in the list we prepared.
After doing that, we should replace all $\sum \omega$ by $\sum
(\omega) -  i n \epsilon$ where $n$ is the number of vertices we combined
together to write down its $\sum \omega$.
 $n$ appears in front of $\epsilon$ since
in combining together $\sum (\omega) - i \epsilon$ from adjacent vertices
we have to sum their $\epsilon$ as well.\footnote{${}^{\dagger}$} {We write
here $- i \epsilon$ instead of $i \epsilon$ because we formulate these
rules as computed in \EFf \ representation. The reader may easily check it
by comparing the $\sum \omega$ given in the rules with the previous
$\sum \omega$ which are computed together with the external vertex.}
The last step here is summing
up over all possible trees keeping in mind that the trees are taken
with plus or minus sign depending on their intrinsic signs
as was explained above.

\noindent
4) The analytical expression corresponding to a given diagram is
the product of what we obtained in steps 1), 2), and 3).

We would like to remark here that even though we obtained our rules
by the method given in the beginning of this section, one can
check in every particular case that the expansion of the probability
density $\exp(-F)$ will give the same result. It is clearly so as one can
show the
equations like \Etime \ are always satisfied if the correlation functions
are computed with the help of our probability density. Even though
the probability distribution possesses not only the four point vertices
but also six and higher order vertices, they in fact conspire to give
only the $\varphi^4$ theory diagrams as the higher order vertices
are the products of the four point ones as is clear from \Elambda,
\Eomega, etc.
If one tries to rigorously prove the Feynman rules by expanding the
probability density, one obtains the expressions which, though much
more complicated than those given above, reduce to  those simpler
expressions after some tedious algebraic computations. It actually looks
like a miracle that so many terms cancel in this procedure. To prove
the algebraic identities appearing in this way is probably a very
interesting but rather difficult task which we have not accomplished but
which is not needed for our purposes.


Let us compute  the expression for the first diagram from the
\Fdiag \ using these rules to see how they work in real life.
The first tree we choose goes along the lines (1, 5). The
expression it corresponds to is
\eqn\Etreeone { {f_1 f_5 \over (\omega_1+\omega_2-\omega_3-\omega_4-2i
\epsilon)
(\omega_5+\omega_6-\omega_3-\omega_4-i \epsilon)}. } We have
used a shorthand notation here $f_n \equiv f_{p_n}$.

Then there is another tree going along the lines (1, 6).
Its expression differs from \Etreeone \ only by replacing $f_5$ by
$f_6$. So we can sum them up together to obtain
\eqn\Etreetwo { {f_1 (f_5+f_6) \over
(\omega_1+\omega_2-\omega_3-\omega_4-2i
\epsilon)
(\omega_5+\omega_6-\omega_3-\omega_4-i \epsilon)
}. }
This is a part of a general rule. If there's a minimal loop consisting
 of the lines $n$ and $m$ going
between two vertices, then we always have the
factor of the form
\eqn\Econt{f_n+f_m  \over \omega_n+
\omega_m-\sum \omega } if the arrows of $n$ and $m$ are in the same
direction and
\eqn\Econter{f_n-f_m  \over \omega_n-
\omega_m-\sum \omega } if they are in the opposite direction
plus some other contributions which can appear
if the tree trunk can go through both ends of the loop we consider  without
going along the lines $n$ or $m$.

Then there are trees going through (2, 5) or (2, 6). Their expressions
are also almost the same as \Etreeone \ with the replacement
$f_1 \leftrightarrow f_2$. We sum them up to obtain
\eqn\Etreethree { {(f_1+f_2)
 (f_5+f_6) \over (\omega_1+\omega_2-\omega_3-\omega_4-2 i
\epsilon)
(\omega_5+\omega_6-\omega_3-\omega_4 -i \epsilon)}. }

Other trees without branches are (3, 5), (3, 6), (4, 5), and (4, 6).
Their sum gives
\eqn\Etreefour {  {(f_3+f_4)
 (f_5+f_6) \over (\omega_1+\omega_2-\omega_3-\omega_4-2 i
\epsilon)
(\omega_1+\omega_2-\omega_5-\omega_6-i \epsilon)}. }

The remaining trees are
with branches. These are
(1, 3), (1, 4), (2, 3), (2, 4). Their expression
can also be found using the Feynman rules. For example, (1, 3) gives
\eqn\Etreefive { - {f_1 f_3 \over (\omega_1+\omega_2-\omega_5-\omega_6-i
\epsilon)
(\omega_5+\omega_6-\omega_3-\omega_4-i \epsilon)} } while all four
trees can be summed up to give
\eqn\Etreesix { - {(f_1+f_2) (f_3+f_4)
 \over
(\omega_1+\omega_2-\omega_5-\omega_6-i
\epsilon)
(\omega_5+\omega_6-\omega_3-\omega_4-i \epsilon)}. } The minus sign
appeared here as one of the lines of the tree trunks here goes the
opposite way to the arrow of that line.

The total expression for the diagram is the sum of \Etreethree,
\Etreefour, and \Etreesix. If we want to compute the correction to the
kinetic equation, we need to take its imaginary part.
We obtain three terms proportional to the three possible delta
functions. For example, the term with $\delta(\omega_3+\omega_4-\omega_1-
\omega_2)$ comes from \Etreethree \ and \Etreefour \ and is proportional to
\eqn\Eppuzyr { \eqalign {
\left( { (f_1+f_2)
 (f_5+f_6) \over \omega_3+\omega_4-\omega_5-\omega_6} +
{(f_3+f_4)
 (f_5+f_6) \over \omega_5+\omega_6-\omega_1-\omega_2} \right) &
\delta (\omega_3+\omega_4-\omega_1-\omega_2) = \cr  =
{ (f_1+f_2-f_3-f_4)(f_5+f_6) \over \omega_3+\omega_4-\omega_5-\omega_6}
& \delta (\omega_3+\omega_4-\omega_1-\omega_2) }
}

It coincides with the first term of \Epuzyr \ after multiplying it
by the necessary $\lambda$'s and by all the propagators $1/f$.

There is an interesting observation applicable for the expressions
we obtain after taking the imaginary part. The $\delta$ functions
which  appear as soon as we take the imaginary part are always
multiplied by the corresponding sum of $f$, that is all of them
appear  in the form $\delta(\sum \omega) \sum f$. It can be proved
by analyzing the rules we discussed or, more easily, by noting that
$f=\omega$ should always be an exact solution of the kinetic equation
\Ekinnexact \ and $\delta(\sum \omega) \sum \omega \equiv 0$ enforces that
last statement.

As we start to compute higher order diagrams, additional feature
we shall discover after taking the imaginary part
is the appearance of the products of the odd
number of $\delta$-functions in them.

A very important property of the diagrams of almost any field theory
which our diagrams are missing is the ``block summing'' property.
Our rules are such that adding a block inside one diagram is totally
different from having this block inside another one.
This poses a major difficulty in dealing with our technique.

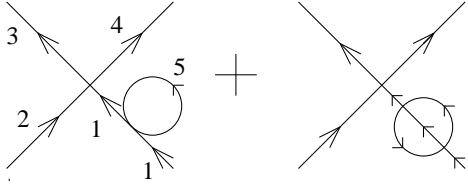
\ifig\Fzdiag{Corrections to the propagators}
{\epsfxsize
2.0in\epsfbox{diag1.ps}}

Now, we immediately ran into new difficulties as soon as we
consider  self energy corrections diagrams like those shown on \Fzdiag.
Such diagrams have at least two lines with the same momentum (because
of the momentum conservation at vertices) and, by using our rules,
we arrive at divergent expressions of the form
$${1 \over \omega - \omega + i \epsilon} = {1 \over i \epsilon} $$
(For example,
the first diagram of \Fzdiag \ has a vertex with a tadpole loop growing
from it and its $\sum (\omega)$ is cleary zero.) It is in fact in the
expressions of this kind that the coefficients in front of $\epsilon$
become important.

The appearance of the divergent expressions is nothing new; we already
dealt with them in the section 2 to discover that they are actually
zero by virtue of the kinetic equation. But there are still  finite
parts left. To compute them, we have to make use of the following
important formulae:
\eqn\Eform {
\int dx {f ( x) \over x \pm i \epsilon} = \sum_{n=0}^{\infty} {(\mp i
\epsilon)^n \over n!} \int dx {f^{(n)}(x)  \over x \pm i 0 }}
\eqn\Evyshe {
\int dx {f(x) \over (x \pm i \epsilon)^n} = {1 \over n!} \int dx {
f^{(n-1)}(x) \over x \pm i \epsilon }}

These formulae  are correct if the integrals
are convergent. If they are not, we have to include the boundary
terms on the right hand side.

Let us use them to calculate the first of two diagrams of \Fzdiag.
We shall discover that this diagram, after adding to it all its
generalizations, including those with many tadpole loops on the
external lines, give the correction to the frequency $\omega$ in
the kinetic equation \Ekin. Really, the first diagram of \Fzdiag \
can be computed to give
\eqn\Eccc {\lambda_{p_1 p_2 p_3 p_4}^2 { f_{p_1}+f_{p_2}-f_{p_3}-f{p_4}
\over f_{p_1} f_{p_2} f_{p_3} f_{p_4} } \delta^\prime ( \omega_{p_1}+
\omega_{p_2}-\omega_{p_3}-\omega_{p_4}) { \lambda_{p_1 p_5 p_1 p_5}
\over f_{p_5} }}
$\delta^\prime$ appears as we have to use \Eform \ and \Evyshe \
in giving sense to the expressions we obtained using the rules discussed
above while the convergence of the integrals in \Ekin \ allows
us not to worry about the boundary terms.

The expression here, when combined with the kinetic equation
\Ekin,   clearly gives  the correction to the frequency in the equation \Ekin \
\eqn\Efc {\omega_p \rightarrow \omega_p + \int
{ \lambda_{p q p q} \over f_q} dq}
While \Eccc \ is just the first term of the expansion due to \Efc \
we can easily prove that
\Efc \ can be obtained up to any order
 if one sums up  the diagrams with all possible
tadpole graphs on the external lines.
However, more complex diagrams, like the second one from the \Fzdiag, can
lead to the corrections which cannot be interpreted that easily.

We would like to conclude this section with saying that the technique
described here is more general than the standard field theory approach of
the $\varphi^4$ theory. In fact, if we take the limit of $f_p \rightarrow
\omega_p$
we shall discover that the prefactor computed according to our
rules goes into $1$ and the whole expression turns into the standard
$\varphi^4$ theory diagram. This should of course be expected as in this
limit $F \rightarrow H$. So our Feynman rules should be treated as
 the ``turbulent'' generalization of the standard field theory rules.
We believe many beautiful results are hidden in this new technique.

\newsec {Epsilon Expansion}

The kinetic equation \Ekin \ is in general very difficult to
solve (that is, to find such $f_p$ that it is satisfied).
However it has long been realized that if both $\lambda$ and $\omega$
are homogeneous functions of momenta, then we can solve that equation
exactly. Namely, following ref. \RZakh \ we choose
\eqn\Edef { \omega_p=p^{\alpha},{\  \
}\lambda_{{\vec p}_1 {\vec p}_2 {\vec p}_3 {\vec p}_4}=
\lambda_0 ( p_1  p_2  p_3 p_4)^{\beta \over 4}
U({\vec p}_1,{\vec p}_2,{\vec p}_3,{\vec p}_4) \delta({\vec p}_1
+{\vec p}_2-{\vec p}_3-{\vec p}_4) }
where  $U$ is a function depending only on
the ratio of lengths of the momenta and the
angles between them and $\lambda_0$ is a small constant.
The parameter $\alpha$ is called the energy spectrum dimension, while
$\beta$ is the interaction dimension.
Then the kinetic equation can be solved with the aid of
the so-called Zakharov transformations, to give
\eqn\Esol {f_p=p^{\gamma}, \gamma={2 \over 3} \beta + d {\rm \ or \ } \gamma=
{2 \over 3} \beta + d - {\alpha \over 3} }
$d$ being the number of space dimensions.
The latter of the solution corresponds to the flux of the wave number while
the former to the energy flux. All that has been known for a long time.

However the solution for $f_p$ we get in this way is just a first
approximation to the full answer which should  in fact be a series in powers
of $\lambda_0$ obtained as a solution to the exact kinetic equation, with all
the higher order corrections.  The situation is reminiscent to that of the
second order phase transition theory (on the theory of the second order
phase transitions, see ref. \RPok \ and references therein). The behavior
of the perturbation series depends crucially on whether the physical
dimension (measured in powers of momentum)
of $\lambda_0$ is positive or negative. Let us denote that dimension
by $-\kappa$. Then, just from the analysis of the dimensions, we see that the
series for $f_p$ should look like
\eqn\Essol {
f_p=\sum_{n=0}^{\infty} c_n \lambda_0^n p^{\gamma+n \kappa} }
$c_n$ being just some dimensionless coefficients. This is of course
true only if the momentum cutoff does not appear explicitly in the
series, that is all the integrals in the kinetic equation are convergent, but
let us first examine this case.
In general we are interested in the behavior of the spectrum $f_p$ at the
distances much larger than the dissipation length. That means that
we should actually take $p \rightarrow 0$ when analyzing \Essol \ (like
in the phase transition theory again). Now if $\kappa > 0 $ then we can
just retain the first term in the series \Essol, while if $\kappa<0$, we
can no longer do that. We must sum up the whole series and analytically
continue it to the region where $p \rightarrow 0$.
So, $\kappa$ plays the role of the $\epsilon=d-4$ parameter of
the theory of the second order phase transitions.

The method of phase transition theory to sum up the \Essol-like series
for $\kappa$ a small negative number is called the ``$\epsilon$-
expansion'' (refs. \RWil \ and \RPok).
We are going to apply this method to the theory we have here.

First let us find what $\kappa$ is equal to. It is clear the physical
dimension of the free wave part of the Hamiltonian \Eham \ and
the interaction part should be the same which leads us to
$$ \alpha-\gamma =  -\kappa+\beta-d+2(-d-\gamma)+4d$$ or
\eqn\Ekappa { \kappa=\beta+d-\gamma-\alpha }
We took into account that the dimension of the fields $a$ and $a^\dagger$
is $-\gamma/2-d/2$.

Let us note that upon substitution of \Esol \ into \Ekappa \ we obtain
$$ \kappa= {\beta \over 3}-\alpha {\rm \ \ or \ \ } \kappa={\beta \over
3} - {2 \over 3} \alpha $$
which shows that actually $\kappa$ does not depend on the
space dimensionality at all. This is different from what we usually have
in the
second order phase transition theory.

Each next correction to the kinetic equation has a higher power of $\lambda_0$.
But the total physical dimension of all the terms should be the same.
That means the total power of the momenta in each next term will be greater
by $\kappa$ than that in the previous one. $\kappa>0$  may lead
to the  divergent integrals in the perturbation series. However
the divergency of the theory we consider does not depend on the power
of the integrand only. It also depends on the properties of the
interaction coefficient $U$. If $U$ decreases exponentially when
the ratio of the momenta becomes high, then all the integrals will
be convergent. In this case the solution $f_p=p^\gamma$ is not
modified at $p \rightarrow 0$
for all $\kappa \ge 0$ (and even exact at $\kappa=0$)
while $\kappa < 0 $ is very difficult, if
not impossible,
to analyze.   That is why, as was promised in the Introduction, we
are going to consider the opposite case, $U=1$, when  the
properties of the interaction do not influence the convergence of the
integrals at all\footnote{${}^{\dagger}$} {We must admit that it is
the intermediate case which is usually fulfilled in Nature, that is
$U$ behaves as a power asymptotically, but we shall consider $U=1$  to
simplify everything as much as we can.}.

Now $\kappa > 0$ does lead to the divergent integrals.
However
it should not bother us  as all the integrals are supposed to
 be cut off at some
high momentum $\Lambda$ where the energy dissipation takes place.
The value of these integrals will depend on the value of the cutoff
only and they change the kinetic equation \Ekin \ only by multiplying it
by a number without changing its form.
The situation is different when $\kappa<0$. Then the integrals in the
perturbation series are more and more convergent and we should compute
the sum \Essol.

   There is also the third possibility, namely $\kappa=0$. In this
case we can expect logarithmically divergent integrals and the series
\Essol \ may turn into
\eqn\Essollog {f_p=p^\gamma \sum_{n=0}^{\infty} c_n \lambda_0^n \log^n \left(
{\Lambda \over p} \right)
 }
The standard approach of the $\epsilon$-expansion theory tells us that
if we are able to separate the most divergent diagrams which contribution are
of the order of $\lambda_0^n \log^n (\Lambda/p) $ and sum them up,
we can obtain a true asymptotic behavior $f_p$ for large logarithms.
Now if $\kappa<0$ we can apply the trick of replacing the logarithms
in \Essollog \  by $p^\kappa$  which will give us a modified wave density
spectrum. This is a program we are going to fulfill.

As was promised many times,  we choose $U=1$ in the interaction, or
\eqn\Ente {
\lambda_{{\vec p}_1 {\vec p}_2 {\vec p}_3 {\vec p}_4}=
\lambda_0 ( p_1  p_2  p_3 p_4)^{\beta \over 4}
 \delta({\vec p}_1
+{\vec p}_2-{\vec p}_3-{\vec p}_4). } Also there will be
an additional condition $\alpha<1$ we will have to impose later  for
the purpose of simplifying the calculations and which is very natural
for the theory described by \Eham \ (see \RZakh).

Let us begin with analyzing the simplest bubble diagrams shown on
\Fdiag.
The special form of the interaction \Ente \ enables us to disentangle the
momenta in it to bring it to the form
\eqn\Eglav {
 -2 K({\vec p}_1, {\vec p}_2, {\vec p}_3, {\vec p}_4) \int dp_5 dp_6
{p_5^\gamma+p_6^\gamma \over p_5^\gamma p_6^\gamma (p_5^\alpha + p_6^\alpha
- p_3^\alpha-p_4^\alpha)} (p_5 p_6)^{\beta \over 2} \delta (
{\vec p}_3+{\vec p}_4-{\vec p}_5-{\vec p}_6 ) }
where we denoted the kernel of kinetic equation by $K$,
\eqn\Ekernel { \eqalign {
K({\vec p}_1, {\vec p}_2, {\vec p}_3, {\vec p}_4) =
{4 \pi \over
f_{p_1} f_{p_2} f_{p_3} f_{p_4} }
& \lambda^2_{
{p_1} p_2 p_3 p_4} (f_{p_1} + f_{p_2} - f_{p_3} - f_{p_4})
\cr & \delta (\omega_{p_1} + \omega_{p_2} -
\omega_{p_3} - \omega_{p_4} )} } and
assumed the integrals over $p_1$, $p_2$, $p_3$ and $p_4$ will be computed
later.

The integral in \Eglav \ will be exactly logarithmically divergent at
large momenta for $\kappa=0$. Namely if $p_5$ and $p_6$ are very
large we can safely
assume ${\vec p}_6=-{\vec p}_5$ eliminating the $\delta$-function in this
way and be left with just one integral giving the $\log(\Lambda)$ at
its upper limit. What is in the lower limit is not important, it is enough
for us to write $$\log\left( {\Lambda \over u({\vec p}_3, {\vec p}_4) }
\right)
$$
where $u$ is some homogeneous function of the first order.
The correction  to the kinetic equation \Ekin \ we thus obtained is
just
\eqn\Ecorrrr {
\sum_{p_1 p_2 p_3 p_4}  - 2 K (p_1, p_2, p_3, p_4)
 \lambda_0 s \log\left(
{\Lambda \over u({\vec p}_3, {\vec p}_4) } \right)) (\delta_{p p_1}+
\delta_{p p_2} - \delta_{p p_3} - \delta_{p p_4}) }
Combining the correction we just obtained with the kinetic equation itself
we arrive at the modified kinetic equation
\eqn\Emod {\sum_{p_1 p_2 p_3 p_4}  K (p_1, p_2, p_3, p_4)(1-2 \lambda_0 s
  \log\left(
{\Lambda \over u({\vec p}_3, {\vec p}_4) } \right)) (\delta_{p p_1}+
\delta_{p p_2} - \delta_{p p_3} - \delta_{p p_4}) }
s being a number coming from the angular integration.

It is not difficult to persuade ourselves that the sum of   the
contributions of the second and the third
terms of \Epuzyr \ will be equal to \Ecorrrr. Really, it is evident
that
   $$ \eqalign{\sum_{p_1 p_2 p_5 p_6 p_3 p_4 }8 \pi \lambda_{p_1 p_2 p_5 p_6}
\lambda_{p_3 p_4 p_5 p_6} \lambda_{p_3 p_4
p_1 p_2}& \times \cr   {(f_{p_1}+f_{p_2}-f_{p_5}-f_{p_6} )
   \delta (\omega_{p_1} + \omega_{p_2} - \omega_{p_5} -
\omega_{p_6}) (f_{p_3}+f_{p_4})   \over f_{p_1} f_{p_2} f_{p_3} f_{p_4}
f_{p_5} f_{p_6} (\omega_{p_3} +
\omega_{p_4} - \omega_{p_5} - \omega_{p_6})} &
[ \delta_{p p_3} + \delta_{p p_4}] =0
  } $$
since the expression above is antisymmetric under the interchange of
$p_1, p_2$ with $p_5, p_6$. On the other hand, the same expression with
$ \delta_{p p_1} + \delta_{p p_2}$ will give us one half of \Ecorrrr.
A similar thing can be said about the third term of \Epuzyr. Together
all of them give us twice the \Ecorrrr.

Unfortunately, these are not the only divergencies associated with this
diagram. It is clear the diagram is going to be divergent for $p_5 \rightarrow
0$ or $p_6 \rightarrow 0$. The degree of the divergence is just
$\beta / 2 - \gamma$, so it is $-\alpha / 2$ for the energy spectrum when,
due to the condition $\kappa=0$, $\beta$ should be equal to $3 \alpha$
or logarithmic for the wave number spectrum when
$\beta=2 \alpha$. This divergence will bring additional correction to the
kinetic equation
\eqn\Eadditional { {q^{\alpha \over 2} \over a^{\alpha \over 2} } {\rm
\ \ or \ \ }
\log(q/a) }
where $a$ is the IR cutoff while q is the momentum of the order of the
external momenta.

However,  if $q$ is sufficiently small, we can neglect the IR term we
just obtained in comparison with the UV one. We admit that this statement
is rather vague. In fact, to have a  satisfying proof, we need to
sum up all the IR corrections corresponding to different diagrams. Most
probably, there is some sort of condensate being formed like in the very well
known case of Bose gas which will cancel the divergencies.
But this has not been shown yet. We intend to devote another paper to the
IR problem. Here we just assume we can neglect all the IR terms.
This simplifies things a great deal as we can neglect the IR corrections
to the frequency as well. The frequency correction \Efc \ is IR divergent
 with the same power as the bubble diagram we considered. We shall
neglect that correction as well while leaving
the task of evaluating the physical effects it  can lead to
to another paper. Acting consistently, we also neglect all the self
energy correction diagrams.

Now we consider the second diagram given by
\eqn\Eoppo {
 -2 K({\vec p}_1, {\vec p}_2, {\vec p}_3, {\vec p}_4) \int dp_5 dp_6
{p_5^\gamma-p_6^\gamma \over p_5^\gamma p_6^\gamma (p_5^\alpha - p_6^\alpha
+ p_4^\alpha-p_2^\alpha)} (p_5 p_6)^{\beta \over 2} \delta (
{\vec p}_4-{\vec p}_2+{\vec p}_5-{\vec p}_6 ) }
It turns out
the condition $\alpha<1$ makes
it convergent. Really, ${\vec p}_5 \approx {\vec p}_6$ in this case,
thus enforcing $p^\alpha_5-p^\alpha_6 << 1$. The power of the integrand
is then
estimated to be $\gamma-1-2 \gamma + \beta +d= -1 + \alpha < 0 $.

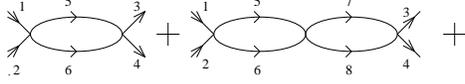
\ifig\Fdiapp{The most important logarithmically divergent diagrams}
{\epsfxsize
2.0in\epsfbox{diag2.ps}}

The nontrivial statement about more complex diagrams is
that when  we go to the higher orders, only the diagrams shown on \Fdiapp \
which are the direct generalization of \Epuzyr \ will give us the necessary
 power of the logarithm.
The expressions corresponding to them can
be easily established by using the rules of the previous section to be
just the product of \Evstav \ for each loop.
In addition to the logarithm to the power of $n$ each of such bubble
diagram
will bring a factor of $(-2s)^n (n+1)$. $n+1$ appears
as the diagrams come with
$2n$ replicas different by the original ones by the interchange of indices
which give us the same expression as the original ones divided by 2.

We can simply sum up the whole series
\eqn\Eprogression{
 \sum_{n=0}^{\infty}(n+1)(-2s \lambda_0)^n \log^n \left( {\Lambda \over
    u({\vec p}_3, {\vec p}_4) } \right) =
{ 1 \over { \left[ { 1 + 2s \lambda_0 \log \left( {\Lambda \over
    u({\vec p}_3, {\vec p}_4)  } \right) } \right] }^2 }  }

So we obtained the full kinetic equation, with the most important
higher order contributions summed up
\eqn\Ekinnfulll {
\eqalign {\int dp_1 dp_2 dp_3 dp_4 &(
\delta_{p p_1}+
\delta_{p p_2} - \delta_{p p_3} - \delta_{p p_4}) \cr
&{\lambda^2_{
{p_1} p_2 p_3 p_4} (f_{p_1} + f_{p_2} - f_{p_3} - f_{p_4})
\delta (\omega_{p_1} + \omega_{p_2} -
\omega_{p_3} - \omega_{p_4} ) \over  { \left[ { 1 + 2s
\lambda_0 \log \left( {\Lambda \over
    u({\vec p}_3, {\vec p}_4)  } \right) } \right] }^2
 f_{p_1} f_{p_2} f_{p_3} f_{p_4}
 } =0 }
}

All of the remaining diagrams will not contribute to the renormalized
kinetic equation because each of the more complex diagram
will have the lines going in the opposite direction and we shall always
encounter expressions like \Eoppo \ while
calculating them.

Let us begin with the integrating out
 the momentum which goes around the minimal loop
(and the minimal loop is the one which goes just between two vertices).
Then we shall obtain the diagram where this loop is replaced by a
dot multiplied by the logarithm of $\Lambda$. Really, the expression
for the diagram came from the sum over different trees. If the
tree went alone one of the propagators of the loop, then the
 the loop momentum should enter the expression in the form of
a multiplier \Econt \ or \Econter \ depending on whether this loop
has  parallel or opposite arrows. The multiplier \Econt \ is
integrated out to give $\log (\Lambda)$ while  the  \Econter \
gives the finite answer.  What is left after the integration is
exactly  the very same diagram with that loop shrunk to a point.
As for the trees not going through the loop,  the integration over
the loop momentum gives the convergent integral as those trees
do not give rise to the inverse loop
propagator $f$ in the numerator (we remember that the $f$'s in the
numerator are taken from the trunk of the tree) thus making the
integral convergent.
If we continue this procedure shrinking the minimal loops one by one,
sooner or later we encounter the loops with the opposite arrows \Econter
\ and the integral will  not acquire the necessary power of the logarithm.
The only exception to this case is the diagrams shown on \Fdiapp.
That is why they are the only diagrams we need to sum up.
Of course, there is an IR problem  left. After integrating over the
minimal loop, we are left with not only the logarithms, but also with the
positive powers of the external momenta due to the IR terms \Eadditional \
(or additional logarithms if the system transports waves instead of
energy). They can lead to  divergent integrals in subsequent
integrations.  For now, we just neglect those IR terms.

Now our task is to find such $f_p$ which solve the equation \Ekinnfulll.
We should
look for $f_p$ in the form of \Essollog.
If we plug it into \Ekinnfulll, we can discover
that
the values of the coefficients $c_n$ depend crucially on the form of the
interaction (except for $c_1$). Fortunately,
it is not the exact form of $f_p$ that we
are looking for. We are interested in the asymptotic behavior of $f_p$ for
large $\log( \Lambda / p)$. This behavior can be shown to be universal.
It is more or less clear even without any computations. \Ekinnfulll \
has an additional $\log^2 (\Lambda)$ in the denominator which should
be canceled by the additional $\log$ in $f$.
That implies that the asymptotic behavior of $f$ is just

\eqn\Efasy {
f_p=p^\gamma \log^{-{2 \over 3}} \left( {\Lambda \over p} \right)
}

The  proof of \Efasy \ is rather messy and not instructive at all, so
we shall not give it while noting that it exists.

We can summarize the result obtained as follows. If $\alpha<1$, the
interaction is chosen in the form (45), and $\kappa=0$, then
 the asymptotic behavior of the
inverse wave density $f_p$ is given by \Efasy.
If $\kappa$ is a small negative number, then the asymptotic behavior
of $f_p$ will be
\eqn\Efasi { f_p=p^{\gamma-{2 \over 3}\kappa } } and the
energy spectrum density will be
\eqn\Eanswer{
 E_p=p^{\alpha - \gamma + {2 \over 3}\kappa} }
The spectrum was modified by ${2 \over 3} \kappa$ which is called the
anomalous dimension.

\newsec {Conclusions and open problems}

So we have shown how one can deal with the turbulence correlation
functions for the system with a small interaction. There is a task
of classifying the IR contributions to the self energy and other
corrections not accomplished here, but it seems the technique we
developed  is powerful enough for this work to be done in the
nearest future.

However, the  most
challenging problem to solve would be to find the correlation functions
for the systems described by the Hamiltonian
\eqn\Ehamh{ H_h =  \sum_{p_1 p_2 p_3 p_4}\lambda_{p_1 p_2 p_3 p_4}
a^\dagger_{p_1} a^\dagger_{p_2} a_{p_3} a_{p_4}
}
 With a particular choice of $\lambda$ this
Hamiltonian describes the motion of the incompressible
fluid (see ref. \RZakh). This theory possesses the integrals $F$ in
the very same fashion as the theory \Eham. Unfortunately we know of
no way to find them explicitly as the equations of motion which follow
from \Ehamh \ are impossible to solve. Still some of their properties
can probably
be studied  by extrapolating the properties of \EFf \ even though the most
pointblank extrapolation, that is $\omega \rightarrow 0$, is most probably
not possible as a phase transition may take place.

It would also be especially interesting to study
these probability distributions in two dimensions.  We may expect
them to give rise to certain conformal field theories with unusual
properties.  The distribution \EFf \ is essentially nonlocal even for
the initially local Hamiltonian while
the actions considered in conformal field theory are usually supposed
to be local.  And while the distribution \EFf \ is explicitly real,
the theory it describes can naturally be nonunitary because of the
possible anomaly in the wave number conservation. The idea of
the conformal invariance of the turbulence correlation functions
was put forward and exploited
in \RConf \ for the case of hydrodynamic turbulence but perhaps
the study of the differences of the standard statistical distributions
and the turbulent distributions can shed more light on this subject.
It is also perhaps possible to formulate a simple statistical
model like Ising model with probability distributions similar to
\EFf \  and study its properties by direct computation.
\vskip 1cm

The author is grateful to A.M. Polyakov for sharing his ideas and especially
for
his support and interest
in this work and to I. Kogan for many
very useful discussions.

\listrefs
\bye